\documentclass[1p,9pt]{elsarticle}

\usepackage[fleqn]{amsmath}
\usepackage{lscape}
\usepackage{geometry}
\usepackage{multirow}
\usepackage{amsmath}
\usepackage{bm}

\usepackage{graphicx}
\usepackage{array}
\usepackage{booktabs}
\usepackage{caption}
\usepackage{subfigure}

\usepackage{array}

\journal{arXiv}

\begin{document}

\begin{frontmatter}

\title{A method for calculating quantile function and its further use for data fitting}

\author{Qing Xiao\corref{cor}}
\ead{xaoshaoying@shu.edu.cn}

\begin{abstract}
This paper introduces a polynomial transformation model based on Weibull distribution, whereby the analytical representation of the quantile function for many probability distributions can be obtained. Firstly, the target random variable $x$ with specified distribution is expressed as a polynomial of a Weibull random variable $z$, the coefficients are conveniently determined by the percentile matching method. Then, substituting $z$ with its quantile function $z=\lambda [-ln(1-u)]^{1/k}$ gives the analytical expression of the quantile function of $x$. Furthermore, using the probability weighted moments matching method, this polynomial transformation model can be used for data fitting. Through numerical experiment, it makes evident that the proposed model is capable of handling some distributions close to binomial which are difficult for the extant approaches, and the quantile functions of various distributions are accurately approximated within the probit range $[10^{-4},1-10^{-4}]$.
\end{abstract}

\end{frontmatter}

\section{Introduction}
The polynomial transformation method is an easy-to-use procedure for simulating various continuous distributions by certain basic probability distribution, such as the standard normal distribution \citep{TPNT2003,TPNT1978,FPNT2002}  and the standard logistic distribution \citep{LmomentLogistic2012}. This technique has been used in various practical settings (see \cite{FPNT2007} paragraph 2).

The polynomial transformation method is expressed as:
\begin{equation} \label{Headrickpolynomial}
  x= \sum_{i=0}^{n}a_iz^i,
\end{equation}
where $x$ is the target random variable, $a_i$ ($i=0,\dots,n$) are undetermined coefficients, $z$ is a standard normal or standard logistic random variable.

The advantage that the logistic variable has over the normal variable is that its inverse cumulative distribution function (ICDF) or quantile function has a closed form $z=ln[u/(1-u)]$. Substituting it into Eq.\eqref{Headrickpolynomial} gives the quantile function of $x$, which is difficult to be obtained analytically for most probability distributions.  Using the associated quantile function, the percentile of $x$ is evaluated directly for a given percentage point, and random numbers with specified distribution can be generated from zero-one standard uniform deviates.

As for the determination of the coefficients $a_i$, the raw moments, central moments and L-moments are employed to perform the moment matching \citep{FPNT2002,LmomentLogistic2012}. Due to the mathematical difficulty, polynomial models of degree $3$ and $5$ are developed, which are limited to a portion of distributions \citep{TPNT2003,FPNT2002}. Yet, this method shows the potential utility for the transformation between different types of probability distributions.

In this paper, the polynomial transformation model is extended to a much higher order (20th) in terms of the Weibull distribution, which also has an analytical representation of ICDF. When establishing the quantile function of the target distribtuion, the percentile matching method is implemented, the coefficients of the model are conveniently assessed by an interpolation method. If the model is used to represent data, the probability weighted moments (PWM) matching method is employed, which requires solving a system of linear equations to determine the coefficients. Finally, numerical examples are provided to check the proposed method.

\section{Calculating the quantile function}
Let $x$ be a continuous random variable, let $z$ be a Weibull random variable with CDF $W(z)$. The distribution $W(z)$ considered is the Weibull distribution with parameters $\lambda=1$ and $k=4$. The quantile function of $W(1,4)$ is:
\begin{equation}\label{InverseCDF}
z=[-ln(1-u)]^{1/4}
\end{equation}
where $u$ is a uniform variable, taking values over the interval $[0, 1]$.

Substitute Eq.\eqref{InverseCDF} into Eq.\eqref{Headrickpolynomial}, the quantile function of $x$ is obtained:
\begin{equation}\label{PolynomialWU}
  x=F^{-1}(u)\simeq a_0+a_1[-ln(1-u)]^{1/4}+\cdots+a_n[-ln(1-u)]^{n/4}
\end{equation}
The parameters in Eq.\eqref{PolynomialWU} can be conveniently evaluated by the percentile matching (PM) method.

The basic idea of PM method is very simple. For a given percentage $u$,  Eq.\eqref{PolynomialWU} should be satisfied. Select $k$ values of percentage: $u_1,u_2,\dots,u_k$, evaluate $x_k=F^{-1}(u_k)$ and  $z_k=[-ln(1-u_k)]^{1/4}$. Using pair values of $(u_k,x_k)$, the polynomial model is obtained by the least square method.

It should be noted that the value of $[-ln(1-u)]^{1/4}$ tends to be positive infinity or 0 with $u$ close to 0 or 1. In this case, the quantile obtained by the proposed model is of high error.

In this paper, a $20$th order polynomial is employed, and $21$ percentage points $u_k$ are chosen evenly over the interval $[10^{-4}, 1-10^{-4}]$. Through numerical example, it is found that accurate results can be obtained for $u\in[10^{-4}, 1-10^{-4}]$.

To assess the proposed polynomial model, comparison is carried out between percentile from the original distribution and those obtained by polynomial model. $10^4$ percentages $u_k$ are chosen evenly over the interval $[10^{-4}, 1-10^{-4}]$. For each percentage $u_k$, the absolute relative error is calculated:
\begin{equation}
\begin{split}
 \varepsilon_k=\Bigg|\frac{x_k^*-x_k}{x_k}\Bigg|\times[\%] ~~~~~x_k=F^{-1}(u_k)~~~~~x_k^{*}=\sum_{i=0}^{n}a_i\left[-ln\left(1-u_k\right)\right]^{-i/4}
\end{split}
\end{equation}
$x_k$ is given by the program of Matlab.

Testing for various probability distributions, the results are shown in Table 1.
\begin{table}[hptb]
\setlength{\abovecaptionskip}{0pt}
\centering
\caption{  The relative error of quantiles for different distributions}
\begin{tabular}[l]{@{}ccc}
\toprule
Distribution	&Parameters	&Maximum value of $ \varepsilon_k$\\
\midrule
Lognormal     &	$0\leq\mu\leq100$  $0<\sigma\leq1$	&$0.07\%$\\
Normal	      &$\mu=0$  $\sigma=1$	            &$0.16\%$\\
Gamma	      &$1\leq a\leq100$  $1\leq b\leq100$	        &$0.05\%$\\
Beta 	      &$1.5\leq a\leq20$  $1.5\leq b\leq20$    	&$0.085\%$\\
Rayleigh	  &$0< \nu\leq200$	                    &$0.0066\%$\\
Chisquare	  &$2\leq \nu\leq100$	                    &$0.075\%$\\
		

\bottomrule
\end{tabular}
\end{table}

Weibull distribution with other parameters is also tried, it is found the Weibull distribution $W(\lambda,k)$ ($\lambda=1$, $3\leq k\leq5$) outperforms others, and $W(1,4)$ stands out for its generality and accuracy.

For the T-distribution ($2\leq \nu\leq100$), Weibull distribution $W(1,6)$ is more preferable, 141 percentage points are are chosen evenly over the interval $[10^{-4}, 1-10^{-4}]$. Using pair values of $(u_k,x_k)$, the polynomial of degree 20 is established by an least square method. The maximum value of $ \varepsilon_k$ is $0.67\%$.

The probability distributions with closed form ICDF have also been attempted for $z$, such as the uniform distribution, exponential distribution and the logistic distribution. Only the logistic distribution yields a good simulation, which is located in the probit range $[10^{-3},1-10^{-3}]$.

\section{Data fitting}
If the PM based polynomial model is used to fit distributions to data, large sample size is required to guarantee a precise value of the percentile $x_k=F^{-1}(u_k)$. Therefore, for a small to moderate sample size, the moment matching method is more preferable.

The statistical information of a random variable is characterized by its statistical moments, a good simulation of the target distribution can be achieved by equating the moments of the polynomial model in Eq.\eqref{Headrickpolynomial} with those of the data. In general, the standardized central moments are involved, the calculation of $r$th raw moments $\left(\sum\limits_{i=0}^{n}a_iz^i\right)^r$ is indispensable, which leads to a complicated computation when the degree of the polynomial is higher than 4 (see \cite{FPNT2002} appendix A). Although the L-moments are employed to simplify the problem, and the analytical expressions of the coefficients are obtained, but the tedious mathematical derivation limits its further use for the polynomial model of a higher degree \citep{LmomentLogistic2011}.

Note that the L-moments are defined as linear combinations of the probability weighted moments (PWM) \citep{LmomentsPWM1990}, mathematically, it is equivalent to perform the moment matching using PWM directly, and the coefficients can be obtained with a much more simple procedure.

The PWM of the random variable $x$ is defined as \citep{PWMY1979}:
\begin{equation}
  M_{p,r,s}=E\left \{x^p\cdot \left[F(x)\right]^r\cdot \left[1-F(x)\right]^s\right\}
\end{equation}
For computational convenience, a particular type of PWM, $\beta_r=M_{1,r,0}$, is considered:
\begin{equation}\label{PWMbeta}
  \beta_r=E\left \{x^p\cdot \left[F(x)\right]^r\right\}=\int_{-\infty }^{+\infty }x\cdot F^r(x)\cdot f(x)dx
\end{equation}
where $f(x)$ is the PDF.

The $\beta_r$ of the polynomial model is:
\begin{equation}\label{PWMequation}
  \begin{split}
  &  \beta_r =\int_{-\infty}^{+\infty}\left(\sum_{i=0}^{n}a_iz^i\right)\cdot W^r(z)\cdot w(z)dz
    =\sum_{i=0}^{n}a_iM^{z}_{r,i,0}\\
     & M^{z}_{r,i,0}=\int_{-\infty}^{+\infty}z^i\cdot W^r(z)\cdot w(z)dz=\int_{0}^{1}[W^{-1}(u)]^i\cdot u^rdu
  \end{split}
\end{equation}
Since $W(z)$ and $w(z)$ are known functions, $M^{z}_{r,i,0}$ can be integrated numerically in terms of the Weibull distribution $W(1,4)$.

The PWM of the observed data is calculated as follows, sort the sample into ascending order $x_1 \leq x_2 \leq \dots \leq x_m$, the unbiased estimate of $\beta_r$ is \citep{PWM2}:
\begin{equation}
 \beta_r=\frac{1}{m}\sum_{s=r+1}^{m}\frac{(s-1)(s-2)\cdots (s-r)}{(m-1)(m-2)\cdots (m-r)}x_s.
\end{equation}

For a polynomial of degree $n$, calculate the first $(n+1)$ PWMs of the data ($r=0,\dots,n$), substitute these PWMs into Eq.\eqref{PWMequation}, the following system of linear equations is established:
\begin{equation}\label{PWMfangc}
\begin{pmatrix} M^{z}_{0,0,0}    & \cdots  & M^{z}_{0,i,0}  & \cdots & M^{z}_{0,n,0} \\
                \vdots           & \cdots  & \vdots         & \cdots & \vdots        \\
                M^{z}_{r,0,0}    & \cdots  & M^{z}_{r,i,0}  & \cdots & M^{z}_{r,n,0} \\
                \vdots           & \cdots  & \vdots         & \cdots & \vdots        \\
                M^{z}_{n,0,0}    & \cdots  & M^{z}_{n,i,0}  & \cdots & M^{z}_{n,n,0}
  \end{pmatrix}
 \cdot \begin{pmatrix} a_0 \\
                \vdots    \\
                a_i \\
                \vdots      \\
                a_n
  \end{pmatrix} =\begin{pmatrix} \beta_0 \\
                \vdots    \\
                \beta_r \\
                \vdots      \\
                \beta_n
  \end{pmatrix}
\end{equation}
Solving the system of linear equation above gives the coefficients of the polynomial model. Then, the PDF is:
\begin{equation}\label{PDF}
f(x)=\frac{1}{[F^{-1}(x)]^{'}}\simeq \frac{1}{\sum^{n}_{i=1}\frac{a_i}{4}\cdot\frac{i}{1-u}\cdot[-ln(1-u)]^{i/4-1}}
\end{equation}

It should be noted that the coefficient matrix in Eq.\eqref{PWMfangc} becomes nearly singular as $n$ increases. Biased values of $a_i$ would be obtained for a polynomial model of a too high degree.

 The generalized lambda distribution (GLD) \citep{GLD1978} and the Johnson system \citep{Johnson1949} are two families of distributions commonly used for data fitting, which are both four-parameter distributions. These two distributions allow for the control of the first four moments of the data, and are often limited to unimodal distributions. While, the proposed method is capable of controlling much higher moments, and can accommodate some distributions that are difficult for the GLD and Johnson system.

\section{Example}
In this section, numerical examples are performed in Matlab to check the proposed method.

As for calculating the quantile function, six examples are performed associated with Lognormal distribution $lnN(0,1)$, standard normal distribution $N(0,1)$, Gamma distribution $\Gamma(10,1)$, Beta distribution $Beta(1.5,1.5)$, Rayleigh distribution $R(1)$ and Chisquare distribution $\chi(3)$. All these distributions are simulated by the PM based polynomial model of degree 20. The PDFs are depicted in Figure 1. The values of $ \varepsilon_k$ are presented in Table 2. Inspection of Figure 1 and Table 2 indicates the accuracy of the PM method.

\begin{table}[!htpb]
\begin{center}
\setlength{\abovecaptionskip}{0pt}
\caption{The absolute relative error}
{\begin{tabular}[l]{@{}cccc}
\toprule
      & Average(\%)  & Minimum(\%)  & Maximum(\%)   \\
\midrule
$lnN(0,1)$ &	$1.8\times 10^{-5}$    &	$0$	&  $0.015$	\\
$N(0,1)$ 	&	$5.1\times 10^{-4}$	   &	$5.1\times 10^{-14}$	&   $0.16$	\\
$\Gamma(10,1)$ 	&	$1.4\times 10^{-4}$    &	$0$	&   $0.074$	\\
$Beta(1.5,1.5)$ &	$1.1\times 10^{-5}$    &	$0$	&  $0.0085$	\\
$R(0.5)$ 	&	$2.1\times 10^{-6}$    &	$0$	&  $0.0015$	\\
$\chi(3)$ 	&	$2.6\times 10^{-5}$    &	$0$	&  $0.030$	\\
\bottomrule
\end{tabular}}
\end{center}
\end{table}

\begin{figure}[!htpb]
\begin{center}
\subfigure[Lognormal distribution $lnN(0,1)$]{
\resizebox*{6cm}{!}{\includegraphics{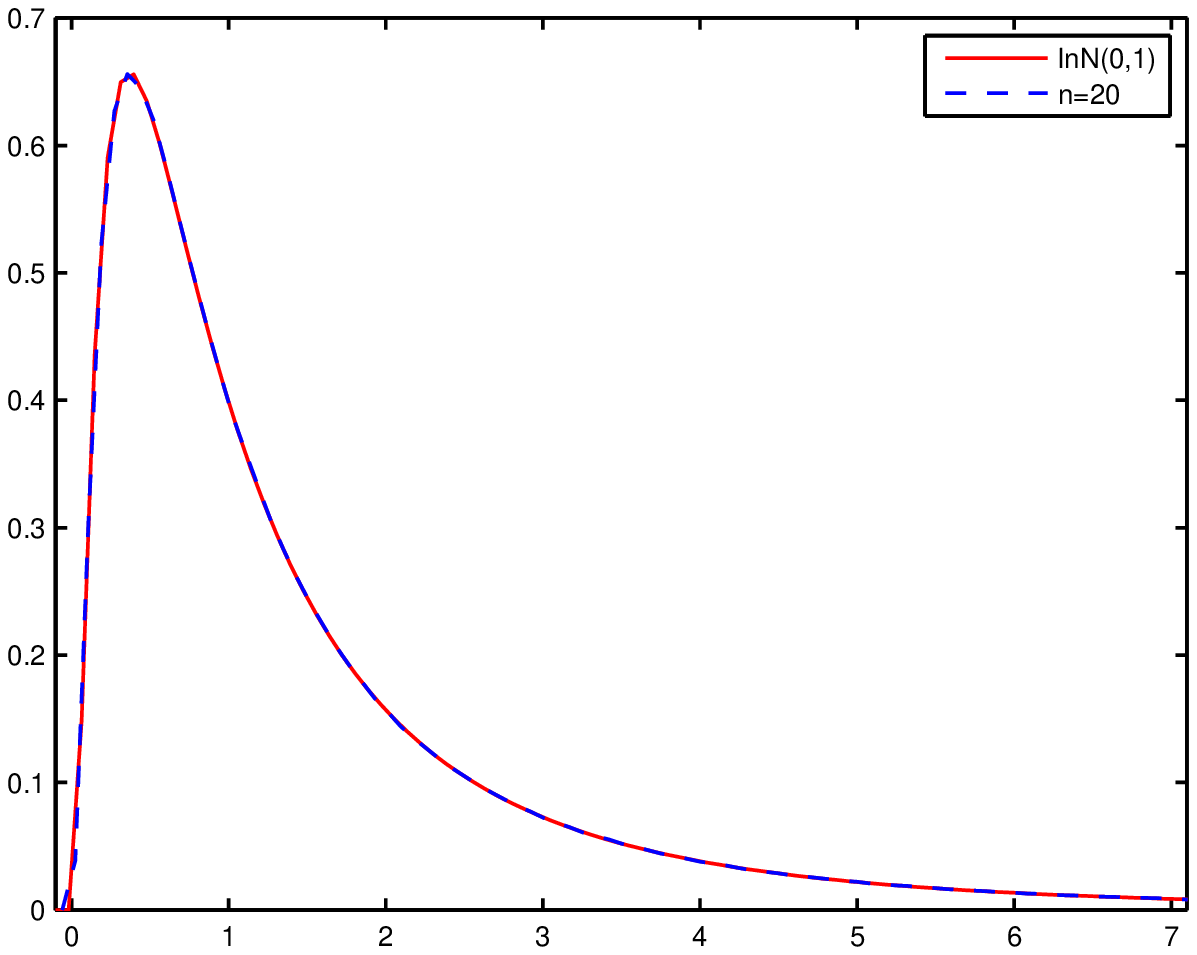}}}%
\subfigure[Normal distribution $N(0,1)$]{
\resizebox*{6cm}{!}{\includegraphics{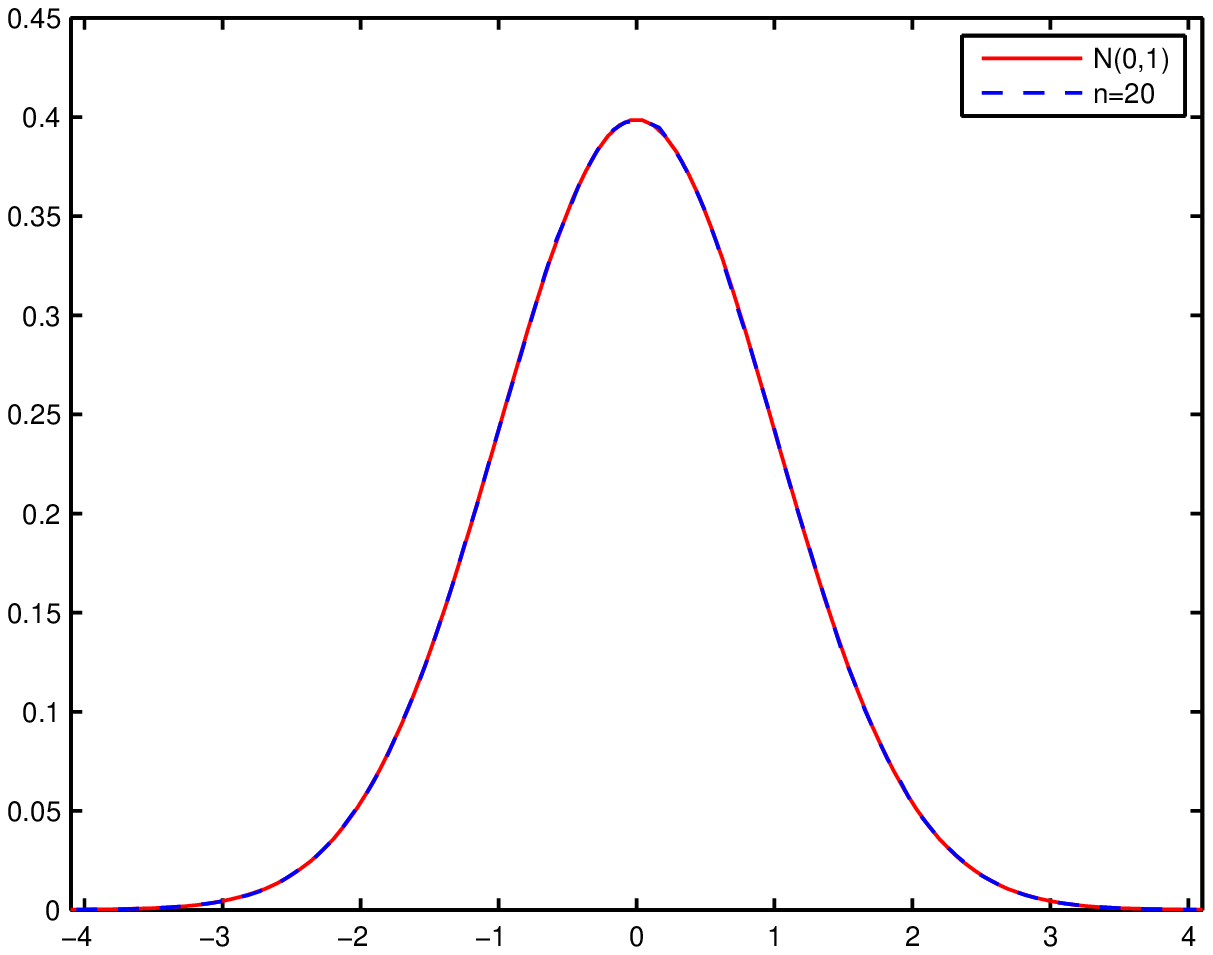}}}%
\\
\subfigure[Gamma distribution $\Gamma(10,1)$]{
\resizebox*{6cm}{!}{\includegraphics{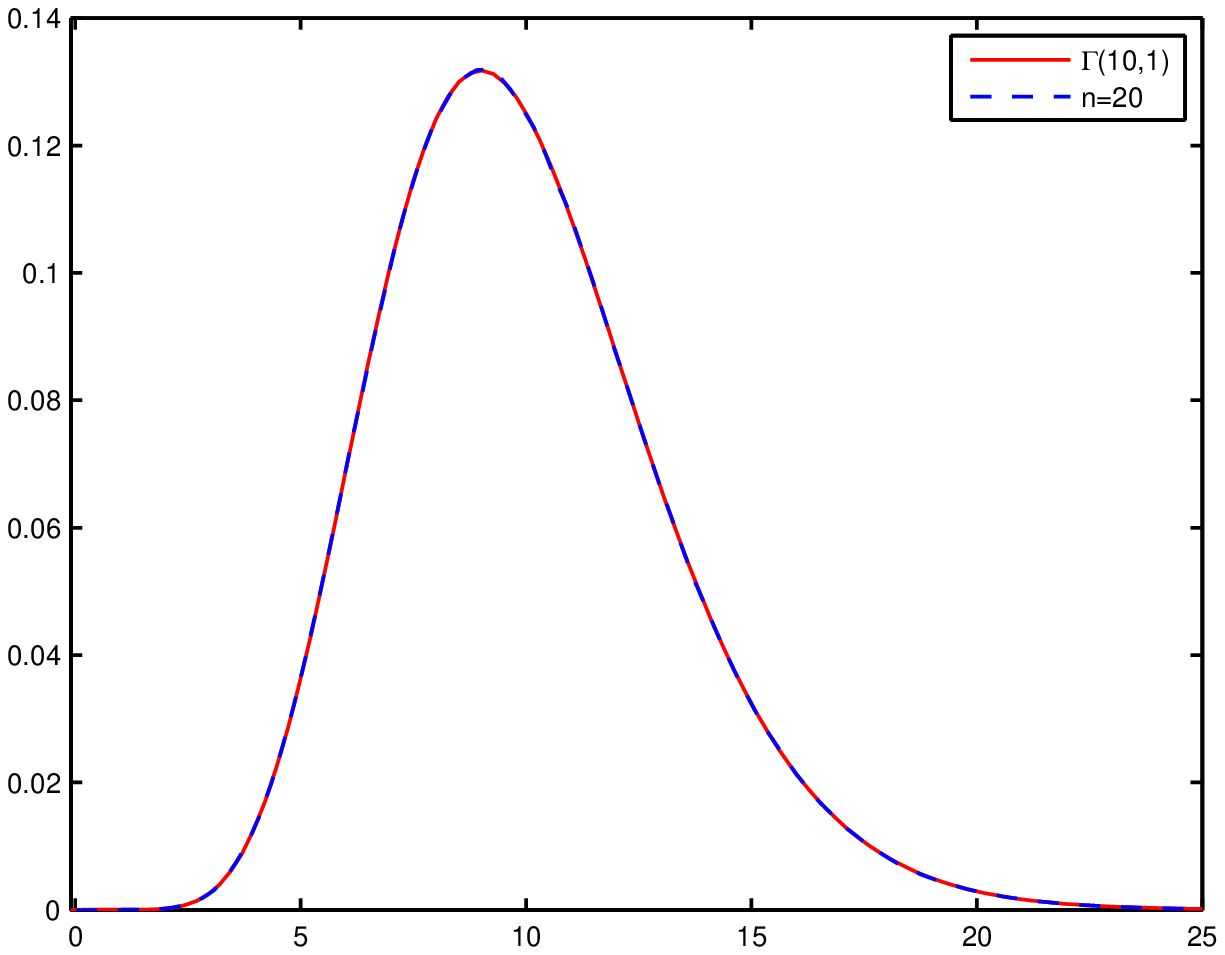}}}%
\subfigure[Beta distribution $Beta(1.5,1.5)$]{
\resizebox*{6cm}{!}{\includegraphics{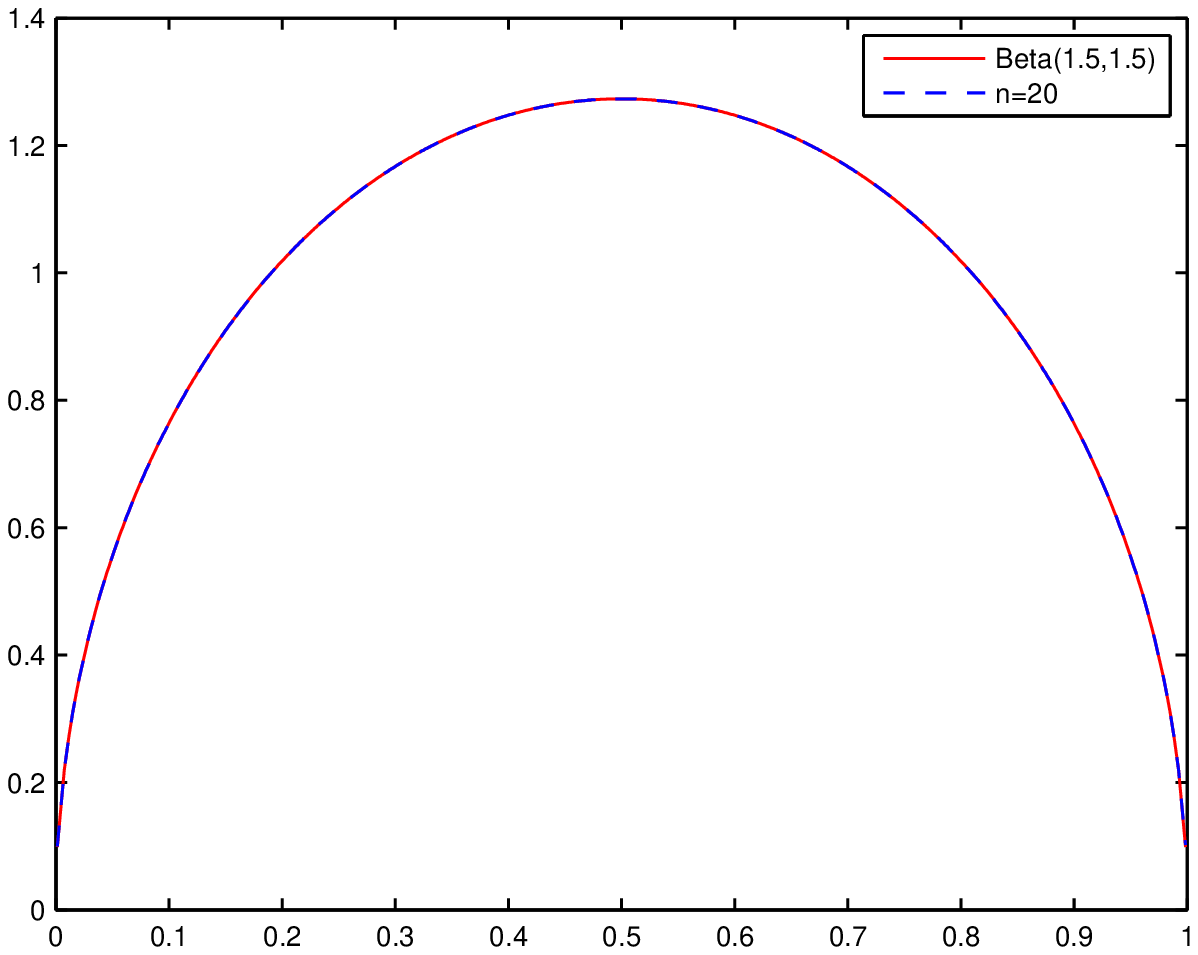}}}%
\\
\subfigure[Rayleigh distribution $R(1)$]{
\resizebox*{6cm}{!}{\includegraphics{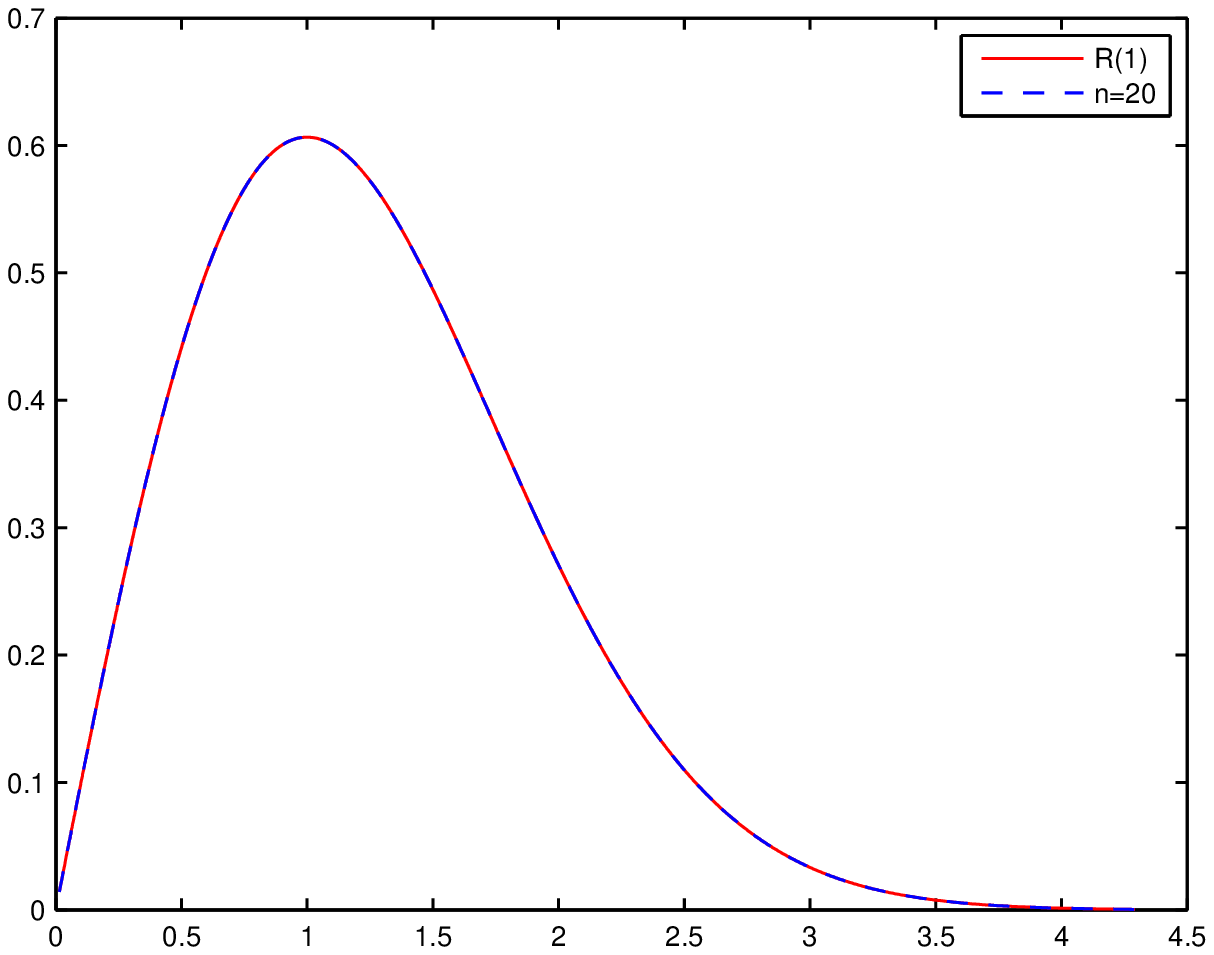}}}%
\subfigure[Chisquare distribution $\chi(3)$]{
\resizebox*{6cm}{!}{\includegraphics{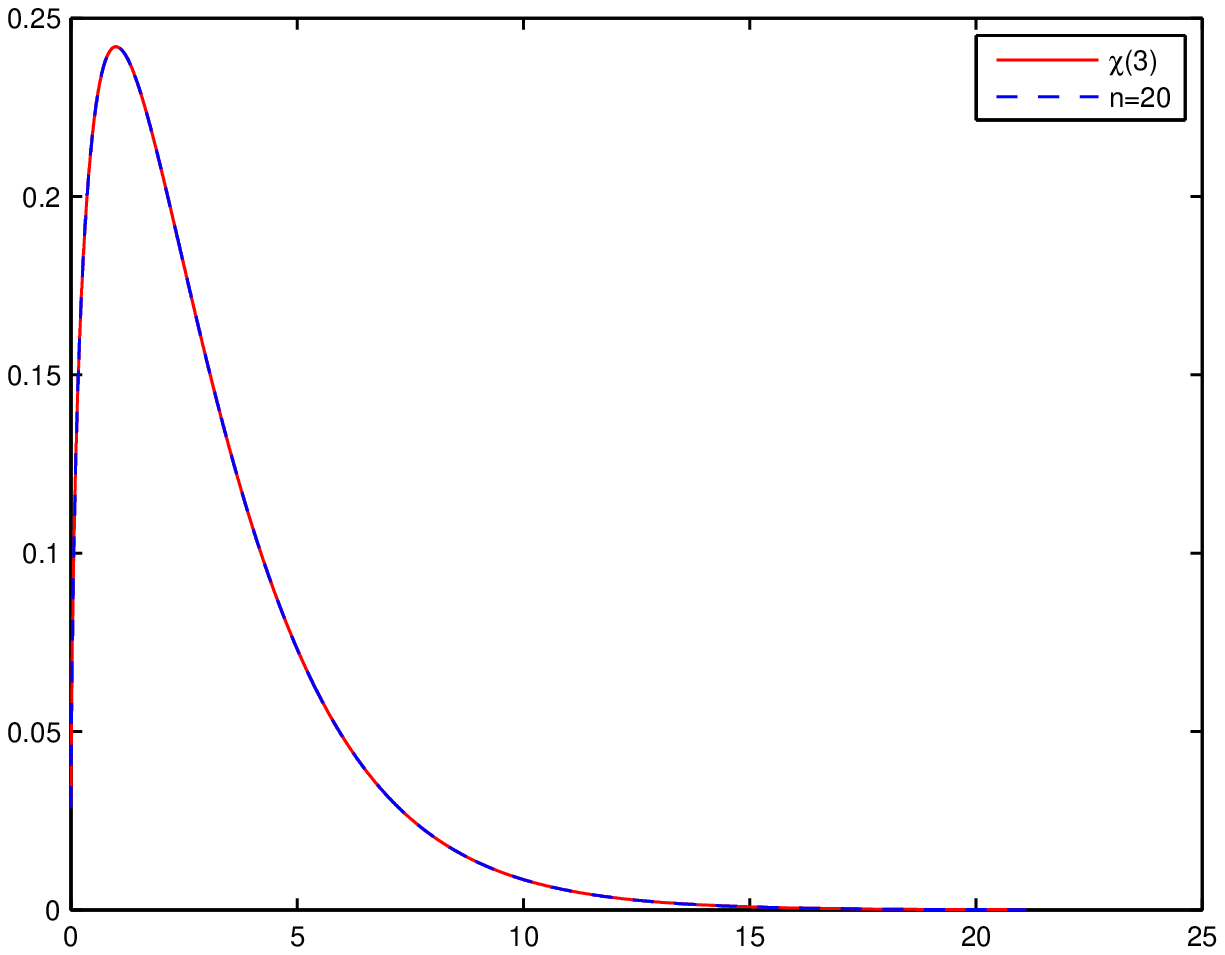}}}%

\caption{\label{fig3} The PDFs of probability distributions}%
\label{sample-figure}
\end{center}
\end{figure}

Here, an example associated with the data fitting is performed.  Consider the random variable $x$:
\begin{equation}
x=x_c+ax_d
\end{equation}
where $x_c$ is a continuous random variable with standard normal distribution $N(0,1)$, $x_d$ is a discrete random variable with binomial distribution $B(1,0.5)$, $a$ is a real constant.
\begin{figure}[!htpb]
\begin{center}
\subfigure[Johnson system fit]{
\resizebox*{5cm}{!}{\includegraphics{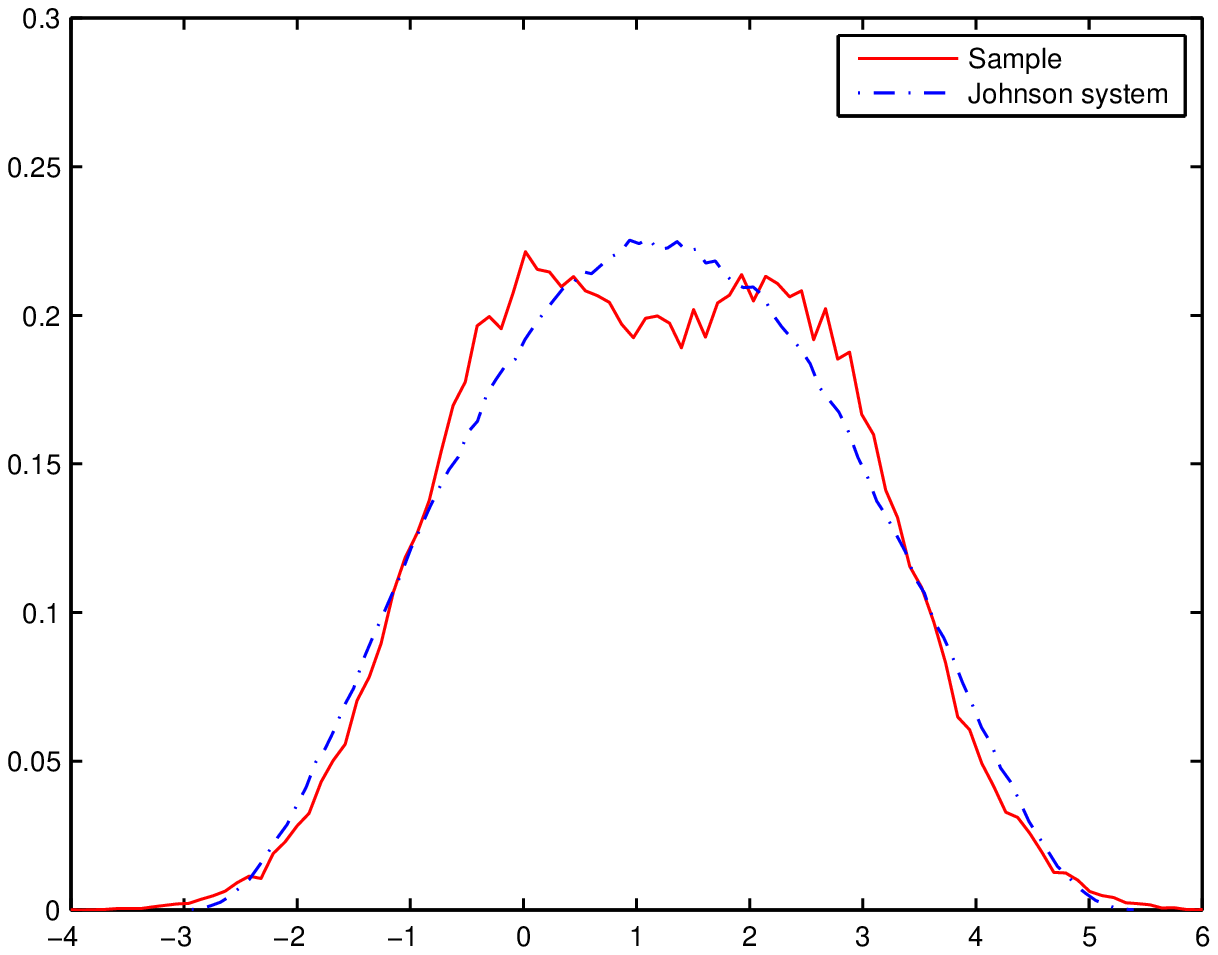}}}%
\subfigure[GLD fit]{
\resizebox*{5cm}{!}{\includegraphics{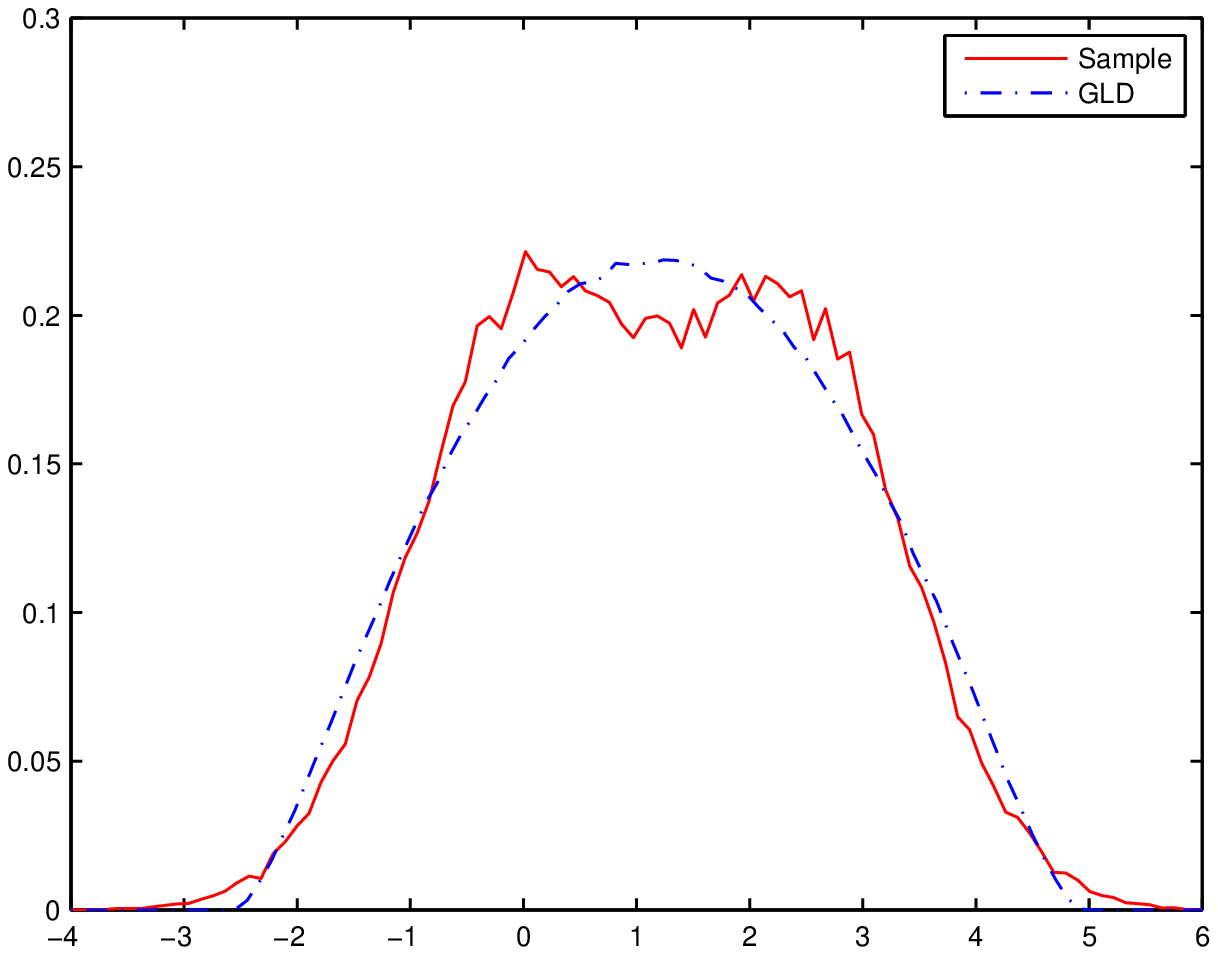}}}%
\subfigure[Polynomial model fit]{
\resizebox*{5cm}{!}{\includegraphics{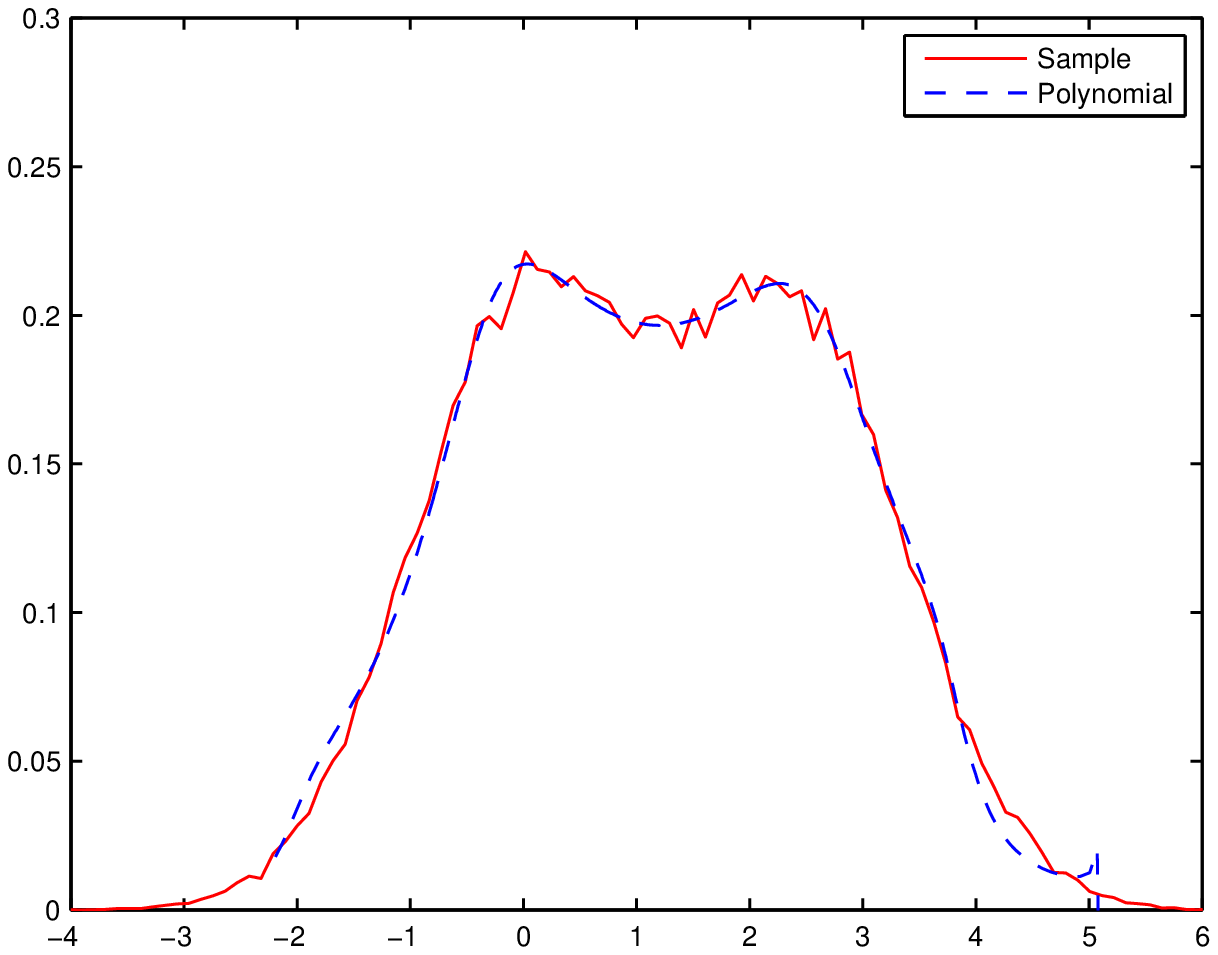}}}%
\caption{\label{fig3} The PDFs the samples}%
\label{sample-figure}
\end{center}
\end{figure}

Setting $a=2.4$, $10^5$ samples are generated, a $20$th order polynomial model is employed, the coefficients are determined by PWM matching method.

The GLD and Johnson system are also employed to represent the data. The parameters of the GLD are estimated by the L-moment matching method \citep{GLDmomentmatchingL}, and the Johnson system is determined by the standard moment matching method \citep{JohnsonM1976}. The PDFs are depicted in Figure 2.

The PDF of $x$ is close to binomial, which is difficult to be simulated by the GLD and Johnson system. While, the proposed model gives a good representation of the samples. In the upper tail part of PDF, the value of $u$ is close to 1, $f(x)$ in Eq.\eqref{PDF} tends to be infinite, leading to the cusp in the right part of the PDF in Figure 2(c).

\section{Conclusion}
A Weibull distribution based polynomial transformation model is proposed in this paper. The quantile functions of various distributions are obtained by percentile matching method, which shows high accuracy in the probit range $[10^{-4},1-10^{-4}]$. When the polynomial model is employed to fit distributions to data, the probability weighted moments matching method is utilized to determine the coefficients. Through numerical examples, it is demonstrated that the proposed model gives a superior simulation as compared to the GLD and Johnson system.

\section*{References}







\end{document}